\documentclass[12pt,twoside]{article}
\usepackage{fleqn,espcrc1}
\usepackage{graphicx}
\usepackage{graphicx}
\usepackage{epsfig}
\newcommand{\AmS}{{\protect\the\textfont2
  A\kern-.1667em\lower.5ex\hbox{M}\kern-.125emS}}

\def\lsim{\raise0.3ex\hbox{$<$\kern-0.75em\raise-1.1ex\hbox{$\sim$}}}
\def\gsim{\raise0.3ex\hbox{$>$\kern-0.75em\raise-1.1ex\hbox{$\sim$}}}

\title{
\vskip -80pt
\mbox{} \hfill ITP Budapest 586\\
\mbox{} \hfill August 2002\\
\vskip 25pt
Lattice QCD Results at Finite Temperature and Density}
\author{Zolt\'an Fodor
\\
\vskip 6pt
Inst. for Theor. Physics, E\"otv\"os University, P\'azm\'any
1, H-1117 Budapest, Hungary
}      
\begin{document}

\maketitle

\begin{abstract}
Recent lattice results on QCD at finite temperatures and
densities are reviewed. Two new and independent techniques
give compatible results for physical quantities. The
phase line separating the hadronic and quark-gluon plasma phases,
the critical endpoint and the equation of state are discussed.
\end{abstract}

\section{Introduction}

QCD at finite temperatures ($T$) and/or chemical potentials 
($\mu$) is of fundamental importance,
since it describes particle physics
in the early universe, in neutron stars and in heavy ion collisions.
According to
the standard picture, at high $T$
and/or high density there is
a transition from a state dominated by hadrons
to a state dominated by partons. 
The expression ``transition''
(which occurs at the transition temperature $T_c$)
is used for first/second order phase transitions and crossovers.
(Observables change rapidly during a crossover,
but no singularities appear.)
Extensive experimental work has been done
with heavy ion collisions at CERN and Brookhaven to explore
the $\mu$-$T$ phase boundary at relatively small $\mu$ values.
At large $\mu$ a rich phase structure is conjectured
\cite{Alford:1997zt,Alford:1998mk,Rapp:1997zu,Rajagopal:2000wf}.

There are well established nonperturbative lattice
techniques to study this
transition at vanishing density. 
At $\mu$=0 we
have fairly good description of the transition (e.g. 
the order of the transition
as a function of the quark masses or the 
the equation of state 
as a function of T).   
For recent reviews see the
summaries of the lattice conferences \cite{U97} or
the review talk on finite T lattice QCD at this 
conference \cite{Kanaya02}. 

\begin{figure*}[htb]
\vskip -0.8truecm
\vspace{9pt}
\hspace*{0.2cm}\epsfig{file=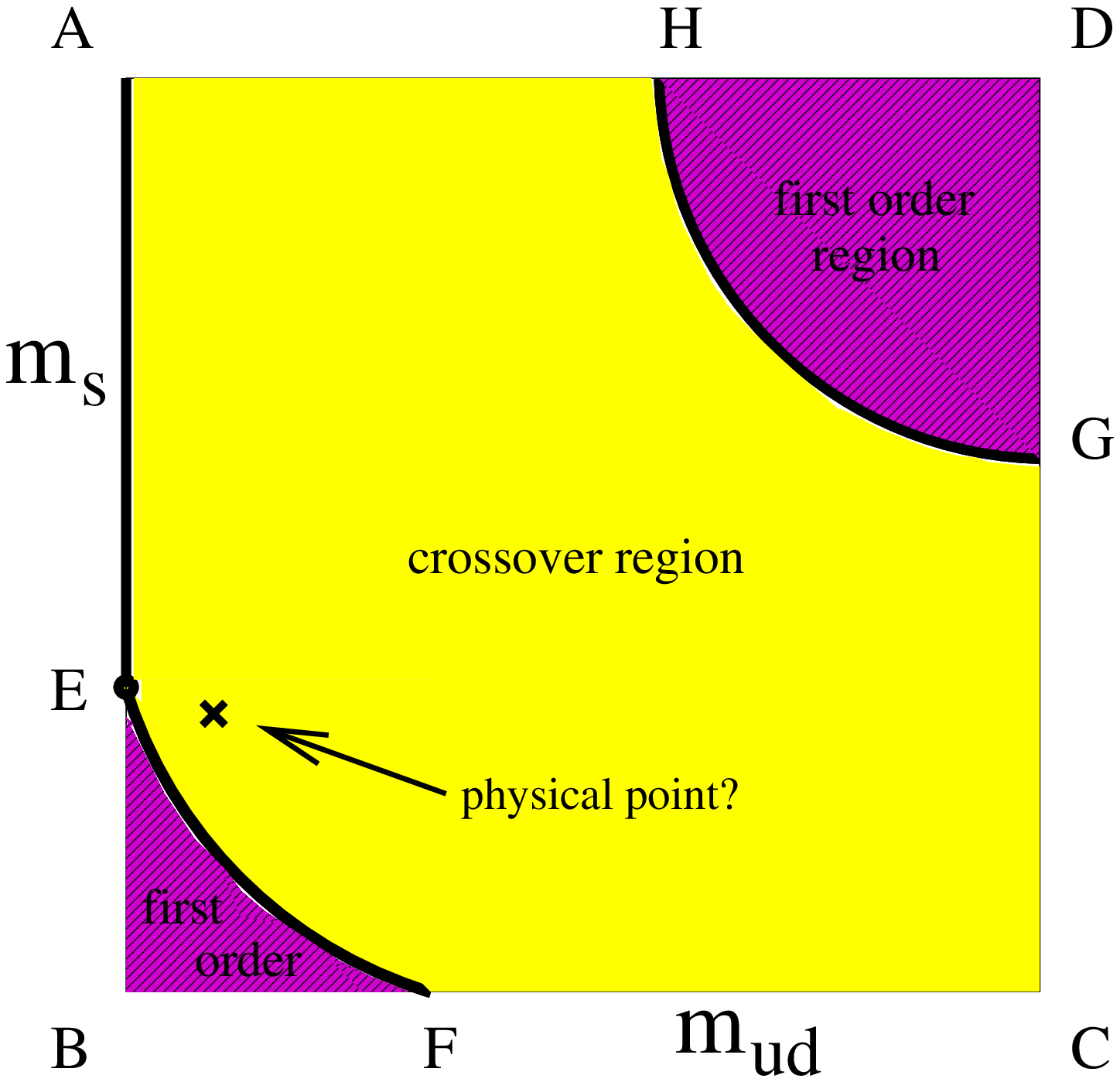,width=59mm}
\hspace*{1.2cm}\epsfig{file=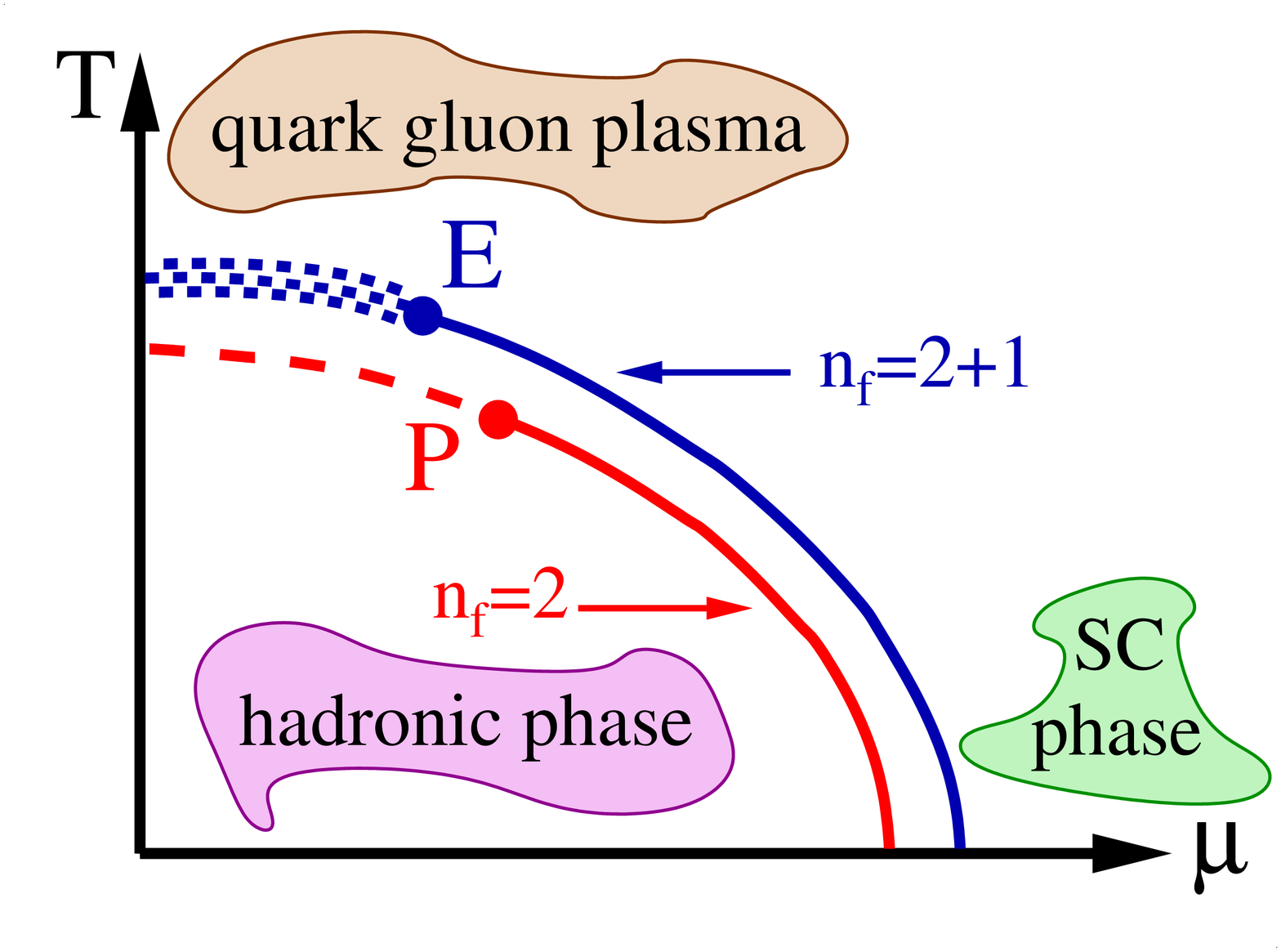,width=81mm}
\vskip -0.7truecm
\caption{
QCD phase diagrams.
Left panel: The phase diagram on the $m_{ud}$ versus $m_s$ plane at $\mu$=0.
The left lower and the right upper corners are first order 
phase transition regions. Thick solid lines 
(A--E,E--F and H--G) 
indicate second order phase transitions. 
The physical point is most 
probably in the crossover region. Right panel: The phase diagram on the 
T versus $\mu$ plane. For small T and $\mu$ the system
is in the hadronic phase, for large T it is a quark-gluon plasma,
whereas for small T and large $\mu$ colour superconductivity
is conjectured. The lower curve shows the $n_f$=2 case.
In the small $\mu$ region the dashed line indicates a second order 
phase transition
which is connected by a tricritical point P to a first order phase 
transition line. The upper curve shows the $n_f$=2+1 case, with 
physical quark masses. The second order phase transition line
disappears, instead of it we are faced with a crossover, illustrated
by the dotted region. This region is connected to a first order 
phase transition line by a critical endpoint E.
}
\vspace{-0.6truecm}
\label{fig:schematic}
\end{figure*}

Our knowledge is far more limited at nonvanishing 
$\mu$. 
Due to the sign problem (oscillating signs lead to cancellation
in results, a phenomenon which appears in many fields of
physics) nothing could have been said for almost 20 years 
about the experimentally important case at nonvanishing
densities. In the last year new, and for the
first time successful approaches
appeared and physically relevant results were obtained 
\cite{Fodor:2001au,Fodor:2001pe,Allton:2002zi,deForcrand:2002ci,DElia:2002aa,Schmidt:2002aa,Fodor:2002aa}. 

The aim of this review is to give a self-contained
picture on the lattice approach at $\mu$$\neq$0 for real QCD
(QCD-like models, such as SU(2), random matrix or NJL models are 
not discussed). 
In Section 2 the qualitative features of
the phase diagram are summarised both at $\mu$=0 and $\mu$$\neq$0. 
Section 3 briefly presents the lattice
formulation and shows the associated sign problem. The main emphasis
is put to the origin of the problem and technical details
are not discussed. In Section 4 the two new 
techniques (the overlap improving multi-parameter reweighting 
\cite{Fodor:2001au} and the analytic continuation 
\cite{deForcrand:2002ci}) are presented. Readers
who are not interested in the origin of the sign problem
and in the new techniques should
skip Section 3 and 4. In Section 5
results of the Budapest group 
\cite{Fodor:2001au,Fodor:2001pe,Fodor:2002aa} 
are listed 
(the phase line, the location of the 
critical endpoint and the equation of state at $\mu$$\neq$0).
They are obtained by the direct application 
of the overlap improving multi-parameter reweighting. 
Section 6  gives the findings 
\cite{Allton:2002zi,Schmidt:2002aa}
of the Bielefeld-Swansea group
(the phase line, the equation of state at $\mu$$\approx$0 and $T$$\approx$$ T_c$
and the response of the critical endpoint to the change of 
$\mu$). They
also use the multi-parameter reweighting technique; however, the 
$\mu$-dependence of the determinant is approximated by a Taylor
expansion. 
Section 7 discusses the phase line obtained by analytic continuation
\cite{deForcrand:2002ci,DElia:2002aa}. Section 8
concludes.

\section{Qualitative features of the phase diagram}

Our knowledge on the phase diagram of QCD at $\mu$=0 and $\mu$$\neq$0 
is summarized on Figure~\ref{fig:schematic}.
Some ingredients are rigorous lattice results, others
are indications from models. 

What are the characteristics of the $\mu$--$T$ phase diagrams, 
relevant for heavy ion collisions?  
(See the right panel of Figure \ref{fig:schematic}.) 
One of the most interesting features of the phase diagram is
a critical endpoint E connecting the first order phase
transition line with the crossover region, 
which separates the low
T hadronic and high T quark-gluon plasma phases. It is 
a long-standing open question, whether such a critical point
exists on the $\mu$-$T$ plane,
and particularly how to predict theoretically its location
\cite{crit_point}.
 
Let us discuss first the $\mu$=0 case (see the left panel
of Figure \ref{fig:schematic}; note, that now we study only 
those features of the figure, which
are relevant for the critical endpoint at $\mu$$\neq$0, for more
details see \cite{Kanaya02} and references therein).
Universal arguments \cite{PW84} and lattice results \cite{U97}
indicate that in a hypothetical QCD
with a strange (s) quark mass ($m_s$) as small as the up (u) and down (d)
quark masses ($m_{u,d}$)
there would be a first order 
phase transition at finite $T$ (point in the E--B--F region). 
The $n_f$=2  case 
with small (but nonvanishing) u/d quark masses but with $m_s=\infty$     
there would be no phase transition only an analytical
crossover. 
This means that between the two extremes there is a
critical strange mass ($m_s^c$) at which one has a second order finite
$T$ phase transition. $n_f$=2+1 lattice results 
with two light quarks and $m_s$ around the $T_c$ 
indicated that $m_s^c$ is about half of the physical $m_s$.
Thus, in the real world we probably have a crossover. (Clearly,
more work is needed to approach the chiral and continuum limits 
\cite{Kanaya02}.) 
 
At nonvanishing $\mu$, arguments
based on a variety of models 
predict a first order finite $T$ phase transition at large $\mu$.
Combining the $\mu=0$ and large $\mu$ informations an interesting
picture emerges on the $\mu$-$T$ plane. For the physical $m_s$   
the first order phase transitions at large $\mu$ should be connected
with the crossover on the $\mu=0$ axis. This suggests
that the phase diagram features a critical endpoint $E$ (with
chemical potential $\mu_E$ and temperature $T_E$), at which
the line of first order phase transitions ($\mu>\mu_E$ and $T<T_E$)
ends \cite{crit_point}. At this point the phase transition is of
second order and long wavelength fluctuations appear, which
results in characteristic experimental consequences, similar to
critical opalescence. Passing close enough to ($\mu_E$,$T_E$)
one expects simultaneous appearance of
signatures, 
which exhibit nonmonotonic dependence on the
control parameters \cite{SRS99},
since one can miss the critical point on either of two sides.
The location of this endpoint is
an unambiguous, nonperturbative prediction of the QCD Lagrangian.
Unfortunately, no
{\it ab initio}, lattice work was done earlier to locate  
the endpoint. Only results from models \cite{crit_point} were available.

The goal of   
present $\mu$$\neq$0 lattice studies is to determine the phase
diagram, to locate the endpoint and also to 
calculate the equation of state.

\section{Lattice QCD and the sign problem at $\mu$$\neq$0}

\begin{figure}[htb]
\vspace{-0.9truecm}
\begin{center}
\epsfig{
file=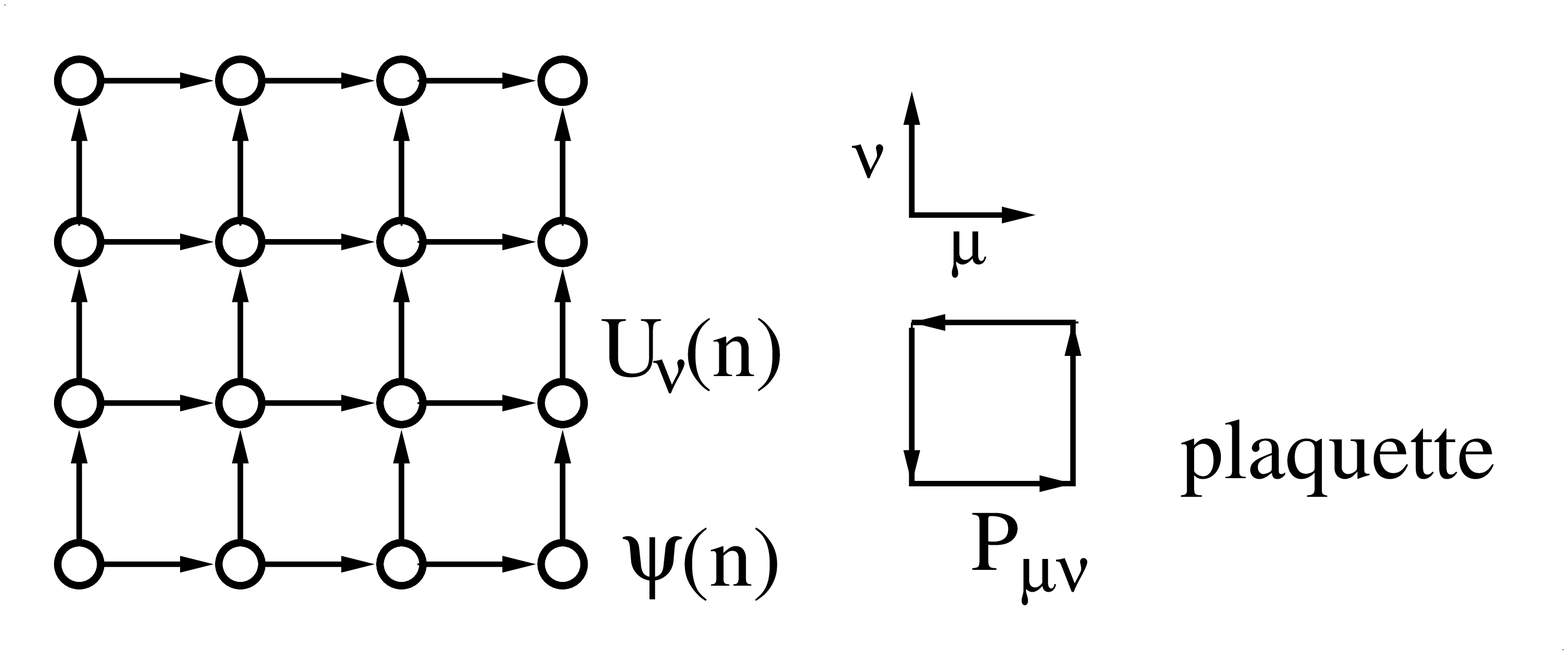,width=100mm}
\end{center}
\vskip -1.6truecm
\caption{
Illustration of Euclidean lattice QCD. Anticommuting quark fields 
are defined on the lattice sites. Gauge fields are SU(3) matrices
and used as link variables: $U_\nu(n)$. 
The product of four links around a 
plaquette leads to the standard gauge action.
}
\vskip -0.6truecm
\label{fig:lattice}
\end{figure}

The continuum QCD Lagrangian in Euclidean spacetime is ${\cal L}=
-{1 \over 4}F_{\mu\nu}^aF^{a\mu\nu}+
\bar\psi(D_\mu\gamma^\mu+m)\psi$. The first term
gives the gauge, the second one the fermionic contribution.
In the discretised lattice formulation the 
anti-commuting $\psi(n)$ quark fields live on the sites ($n$) of the lattice,
whereas the $A_\mu^a$ gluon fields are used as link ($U_\mu(n)$) and as 
plaquette ($P_{\mu\nu}(n)$) variables
\begin{equation}
U_\mu(n)=
\exp{(ig_s\int_n^{n+e_\mu}d{x'}^{\mu}\,A_{\mu}^{a}(x')\lambda_a/2)},
\ \  P_{\mu\nu}(n)=
U_{\mu}(n) U_{\nu}(n+e_\mu) U^{\dagger}_{\mu}(n+e_\nu) U^{\dagger}_{\nu}(n)
\end{equation}
Here $\mu$ represents the direction and $e_\mu$ the corresponding unit 
vector. 
Similarly to the continuum formulation the action 
$S=S_g+S_f$ consists of the pure gluonic and the fermionic parts. The 
gluonic part is written with the help of the plaquettes:
$S_g=6/g_s^2\cdot\sum_{n,\mu,\nu}\left[
1-{\rm Re}({\rm Tr}P_{\mu\nu}(n))\right]$, where $g_s$ is the gauge coupling. 
The fermionic part on the lattice needs a       
differencing scheme for quarks. In the naive, noninteractive
case one obtains: 
$\bar\psi(x)\gamma^\mu\partial_\mu\psi(x)\rightarrow
\bar\psi(n)\gamma^\mu(\psi(n+e_\mu)-\psi(n-e_\mu))$.
The interacting case has $D_\mu$ instead of $\partial_\mu$, thus
the gauge field is also included:
$\bar\psi(x)\gamma^\mu D_\mu\psi(x)\rightarrow
\bar\psi(n)\gamma^\mu U_\mu(n)\psi(n+e_\mu)+...$ (fermion doublers
do not play any role in the sign problem, therefore we do not discuss them).
Fermion fields appear only in bilinear expressions, thus 
we can write
$S_f=\bar\psi(n)M_{nm}\psi(m)$, where M is the $U$-dependent
fermion matrix. Note, that 
the number of raws or coloumns of M is proportional to the lattice
volume. In our case the matrix M is sparse, only diagonal ($\propto m$) 
and next-neighbour ($\propto U_\mu$)
elements are nonvanishing. 

Our system can be 
described by the  
Euclidean partition function. It is given by integrating the Boltzmann 
weights over the gauge and fermionic fields. The action is 
bilinear in the fermionic variables, thus this part of the integral can be 
calculated explicitely.
\begin{equation}\label{eq:part_func}
{\mbox Z=}\int
{\cal D}U{\cal D}\bar\psi {\cal D}\psi e^{-S_g-S_f}
{\mbox =\int}{\cal D}U e^{-S_g}\det M(U).
\end{equation}
The remaining integral over the $U$ fields is calculated
by stochastic methods. 
A canonical ensemble of field configurations is generated by
Monte-Carlo algorithms. Observables are obtained as averages
over the field configurations.
The intrinsic feature of these techniques is 
importance sampling, thus we sample only the most important configurations
and these individual configurations 
have equal weights in the averages.

We use the most straightforward technique, 
the Metropolis algorithm as an illustration (the Metropolis
method is very CPU-intensive, thus usually faster but more complicated
techniques are used).  
The basic step of the method is a stochastic, reversible 
modification of a link variable: $U_\mu(n)\rightarrow U'_\mu(n)$.
It is accepted or rejected.
The Metropolis condition  
\begin{equation}\label{eq:metropolis} 
P(U\rightarrow U')=\min\left
[1,\exp({S_g(U)-S_g(U')})\det M(U')/\det M(U)\right]
\end{equation}
gives the probability to accept the new $U'$ field variable.
It is easy to show, that by sweeping through the whole lattice
many times, we reach the canonical distribution, and physical 
observables can be determined. Note, that for real gauge fields
$\det M(U)$ is real, thus eq.~(\ref{eq:metropolis}) really has 
a probability interpretation. This feature is essential
not only for the Metropolis algorithm, but for any importance
sampling based method.

Our primary goal is to understand QCD at finite densities. As usual,
finite densities can be studied by introducing a chemical
potential. One adds the following term to the continuum action:  
$\mu {\bar \psi}(x)\gamma_4\psi(x)$. In this
formula $\mu$ acts as a 
fourth component of an imaginary, constant vector potential 
\cite{Hasenfratz:1983ba,Kogut:1983ia}.
As we have seen real gauge fields result in real $\det M(U)$;
however, the inclusion of $\mu$ (thus a constant imaginary 
gauge field) gives complex $\det M(U)$. This spoils the
probability interpretation of eq.~(\ref{eq:metropolis}) and
any other importance sampling based method. 
Instead of an ensemble of equally important configurations 
(importance sampling) we might have configurations with
complex Boltzmann weights. These complex weights with oscillating
real parts largely cancel each other in observables. In
the literature this phenomenon
is referred to as the ``sign problem''. 
 
\section{New lattice techniques at finite chemical potential}

\begin{figure}[htb]
\vspace{-1.2truecm}
\begin{center}
\epsfig{
file=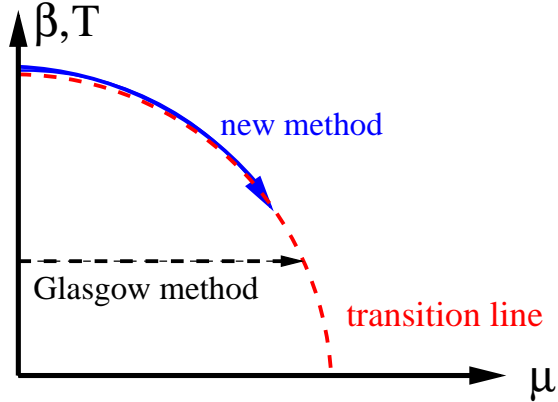,width=80mm}
\end{center}
\vskip -1.6truecm
\caption{
Illustration of the multi-parameter reweighting. The horizontal arrow
shows the Glasgow, the curved one the new, multi-parameter  
reweighting. The dashed transition line separates the different phases. 
The Glasgow method
starts with an ensemble in the low T, hadronic phase
and uses the weighting factors of the configurations
at the same T but at $\mu$$\neq$0. Clearly, the hadronic 
configurations do dot overlap with the  
transition ones.
The new method starts
e.g. at the transition point at $\mu$=0 and changes T and $\mu$
simultaneously, keeping the system on the transition line.
The overlap is much better.
}
\vskip -0.6truecm
\label{fig:method1}
\end{figure}

The overlap improving
multi-parameter reweighting \cite{Fodor:2001au} opened the possibility
to study lattice QCD
at nonzero $T$ and $\mu$. First one produces
an ensemble of QCD configurations at $\mu$=0 and at T$\neq$0.
Then the Ferrenberg-Swendsen type reweighting factors
\cite{Ferrenberg:yz} of these configurations are determined
at $\mu$$\neq 0$ and at a lowered T. The idea can be
easily expressed in terms of the partition function
\begin{eqnarray}\label{eq:reweight}
&&Z(\mu,\beta) = \int {\cal D}U{\rm e}^{-S_{g}(\beta,U)}\det M(\mu,m,U)=
 \\
&&\int{{\cal D}U {\rm e}^{-S_{g}(\beta_0,U)}\det M(\mu=0,m,U)}
{ \left\{{\rm e}^{-S_{g}(\beta,U)+S_{g}(\beta_0,U)}
\frac{\det M(\mu,m,U)}{\det M(\mu=0,m,U)}\right\} },
\nonumber 
\end{eqnarray}
where $S_g(U)$ is the action of the gluonic field,
$\beta=6/g^2$ fixes the coupling of the strong interactions ($g$).
Note that for a given lattice $T$
is an increasing function of $\beta$. The quark mass parameter is $m$ and       
$\det M$ comes from the integration over the quark fields (see eq.(\ref{eq:part_func})). At nonzero
$\mu$ one gets a complex $\det M$ which has no probability interpretation, thus
it spoils any importance sampling.
Thus, the first line of eq. (\ref{eq:reweight}) at $\mu$$\neq$0,
is rewritten in a way that the first part of the second line 
is used as an integration measure (at $\mu$=0, for which
importance sampling works) and the remaining part in the curly
bracket is measured on each configuration and interpreted
as a weight factor $\{ w(\beta,\mu,m,U)\}$. (For $n_f$$\neq$4
fractional
powers of $\det M$ is needed. This complication can be solved 
\cite{Fodor:2001pe}.)

The reweighting is performed along the best weight lines on the
$\mu$--$\beta$ plane (or equivalently on the $\mu$--$T$ plane).
One such line is the transition line.
The best weight lines are determined by minimising the spread of $\log w$.
The technique works for
$T$ at, below and above 
$T_c$.   Using the above weights any observable ${\cal O}$
can be determined at $\mu$$\neq$0  
\begin{equation}\label{eq:opexp}
{\overline {\cal O}}(\beta,\mu,m)=\frac{\sum \{w(\beta,\mu,m,U)\}
{\cal O}(\beta,\mu,m,U)}{\sum
\{w(\beta,\mu,m,U)\}}.
\end{equation}                      
 
The method is illustrated on Fig. \ref{fig:method1}.
Using multi-parameter reweighting one simultaneously changes 
T and $\mu$.
A much better overlap can be obtained by the multi-parameter
reweighting than by the single $\mu$-reweighting Glasgow-method
\cite{Barbour:1997ej}.
The Glasgow-method reweights pure hadronic configurations to 
transition ones. 
By the new technique transition (or hadronic/QGP)
configurations are reweighted to transition
(or hadronic/QGP) ones. 
Since the original
ensemble is collected at $\mu$=0 one does not expect that 
even the new technique
is able to
describe the physics of the large $\mu$ region with e.g. 
colour superconductivity. Fortunately, the typical $\mu$ values
at present heavy ion accelerators are smaller than the covered region.

An alternative approach \cite{Allton:2002zi} uses  Taylor expansion
of eqs.~(\ref{eq:reweight},\ref{eq:opexp}) 
as a function of $\mu$ (or $m$) and hence
estimates the derivatives of various quantities with respect to $\mu$ (or
$m$). Thus, for
the case of a Taylor expansion in $\mu$ one has  
\begin{eqnarray}
\label{eq:expand}
\ln\left({{\det M(\mu)}\over{\det M(0)}}\right) =
\sum_{n=1}^{\infty} \frac{\mu^n}{n!}
\frac{\partial^n \ln \det M(0)}{\partial \mu^n}\equiv\sum_{n=1}^\infty
{R}_n\mu^n.
\end{eqnarray}              
Instead of using the explicit form of the determinants in 
eq.~(\ref{eq:reweight}) the 
Bielefeld-Swansea group used the derivatives 
in $\mu$. Compared to the explicit
calculation of the determinants, the Taylor technique needs less CPU-time
(derivatives can be estimated stochastically); however, 
only valid for somewhat smaller $\mu$ values than the full technique.   
 
The other promising and absolutely independent 
approach is the analytic continuation
from imaginary to real chemical potentials \cite{deForcrand:2002ci}.
One computes first the critical line for 
imaginary chemical potential. For these $\mu$
values there is no sign problem, therefore 
Monte-Carlo simulations based on importance sampling
can be carried out. 
One can check the convergence of the Taylor expansion
of the critical line, thus $\beta_c$ as a function of Im($\mu$).  
The convergence seems to be
surprisingly fast for the whole range of 
chemical potentials accessible to this method.
Analytic continuation of the Taylor series then reduces to simply 
flipping the sign of the appropriate terms.
Finally, the infinite volume limit has to be 
taken from the continued results
at real $\mu$. 
Using this technique
the phase line separating the hadronic and quark-gluon plasma
phases were determined \cite{deForcrand:2002ci,DElia:2002aa}.
The order of the transition can 
then be determined in a $(V,\mu)$-range where
the truncation error of the series is smaller 
than finite size scaling effects. 

\section{Overlap improving multi-parameter reweighting: 
direct approach}

This section summarizes some results of the Budapest group
\cite{Fodor:2001au,Fodor:2001pe,Fodor:2002aa} obtained 
in 2+1 flavour dynamical staggered QCD  on $N_t$=4 lattices. 

\begin{figure*}[htb]
\vskip -1.3truecm
\vspace{9pt}
\hspace*{-0.2cm}
\begin{center}
\epsfig{file=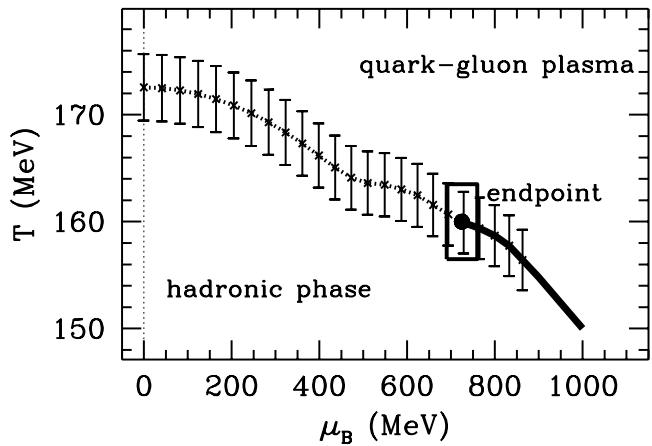,width=90mm}  
\end{center}
\vskip -1.2truecm
\caption{The T-$\mu$ diagram. Direct results are given with errorbars.
Dotted line at small $\mu$ shows the crossover, solid line 
at larger $\mu$ the first order
transition. The box gives the uncertainties of the endpoint.}
\label{fig:phase_diag}
\vskip -0.6truecm
\end{figure*}   

Figure \ref{fig:phase_diag} shows the phase diagram on the $\mu$--T
plane in
physical units, thus $T_c$
as a function of $\mu_B$, the baryonic chemical potential
(which is three times larger then the quark chemical potential).
The analysis is consisted of three steps. First one determines
the transition points as a function of $\mu$. Then 
the $V\rightarrow\infty$ behaviour is inspected to separate
the crossover and the first order phase transition region. 
Finally one transforms lattice units into physical ones.
In physical units the endpoint \cite{Fodor:2001pe}
is at $T_E=160 \pm 3.5$~MeV, $\mu_E=725 \pm 35$~MeV.
At $\mu_B$=0 we obtained $T_c=172 \pm 3$~MeV.   
Note, that due to CPU-limitations
in this analysis the light quark masses are approximately
three times larger than their physical value and the lattice spacing
is $\approx$0.28~fm. Clearly more work is needed to extrapolate
to the thermodynamic, chiral and continuum limits. 

\begin{figure*}[htb]
\vskip -.7truecm
\vspace{9pt}
\hspace*{-0.2cm}\epsfig{file=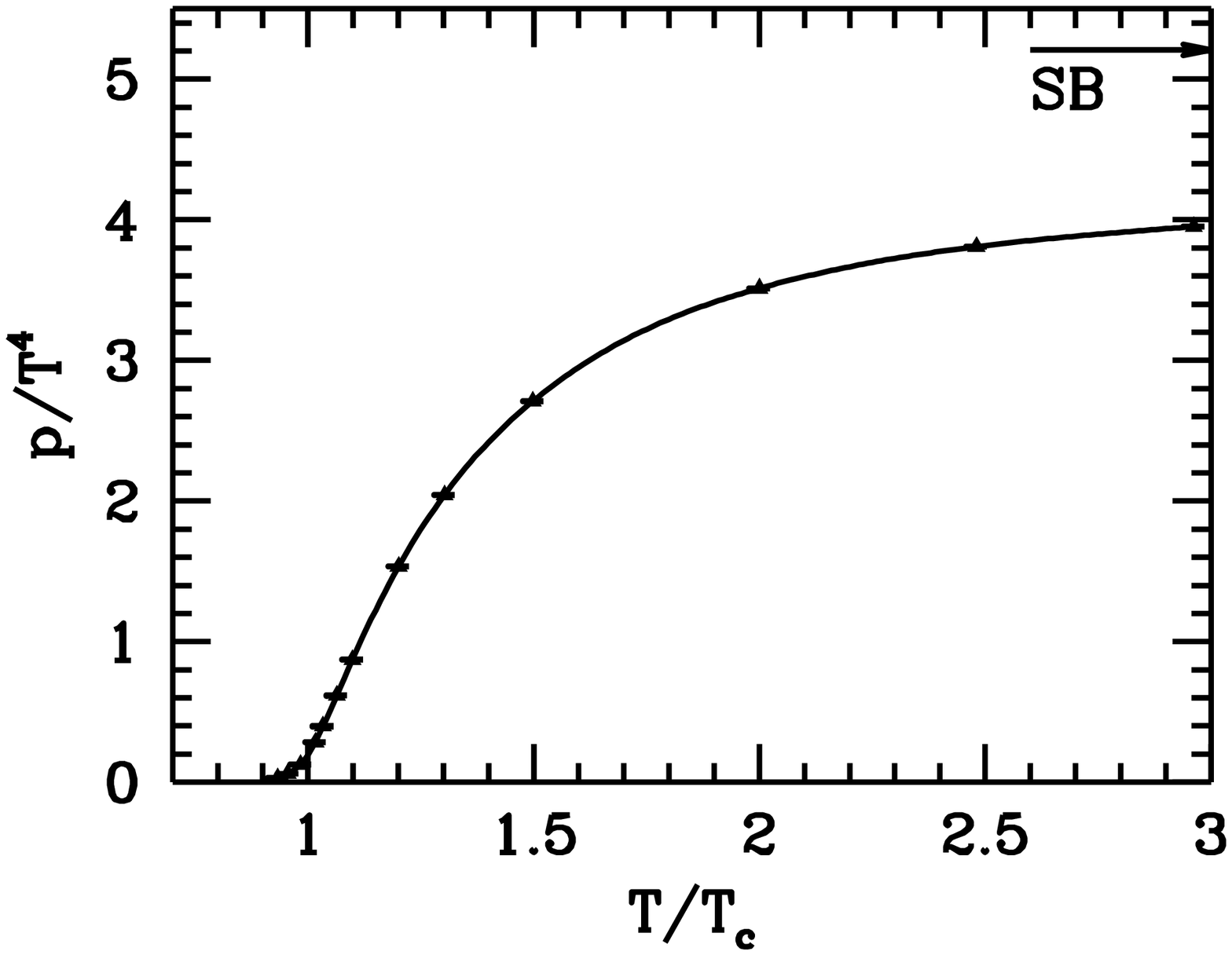,width=80mm}
\hspace*{-0.2cm}\epsfig{file=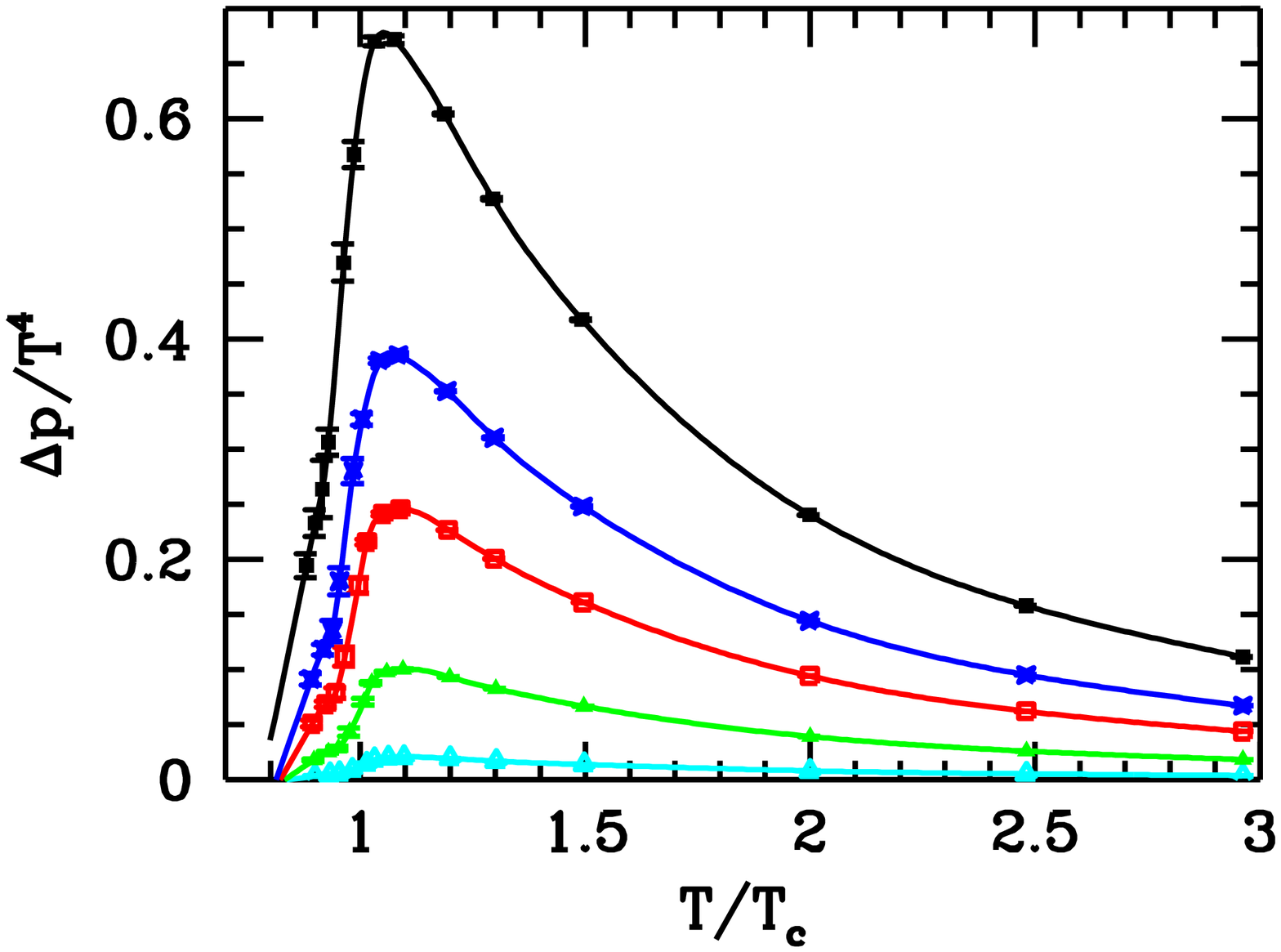,width=80mm}
\begin{picture}(17,0.1)
\end{picture}
\vskip -3.3truecm
\caption{
Left panel: 
$p/T^4$ as a function of $T/T_c$ at $\mu=0$
The continuum SB limit is also shown.
Right panel: $\Delta p=p(\mu\neq 0,T)-p(\mu=0,T)$ normalized by $T^4$
as a function of $T/T_c$
for $\mu_B$=100,210,330,410~MeV and
$\mu_B$=530~MeV (from bottom to top).
}
\vspace{-0.6truecm}
\label{fig:pressure}
\end{figure*}

The equation of state at $T\neq 0$ and $\mu\neq 0$ is also determined
\cite{Fodor:2002aa}. In order to help the continuum 
interpretation of the figures we normalize the raw lattice results with
the dominant correction factors between $N_t$=4 and the continuum
in the $T\rightarrow\infty$ (Stefan-Boltzmann; SB) case 
(see Ref. \cite{Fodor:2002aa}).
Therefore, the results presented on the
figures might be interpreted as continuum estimates and could be directly 
used in phenomenological applications.

\begin{figure*}[htb]
\vskip -1.1truecm
\vspace{9pt}
\hspace*{-0.2cm}\epsfig{file=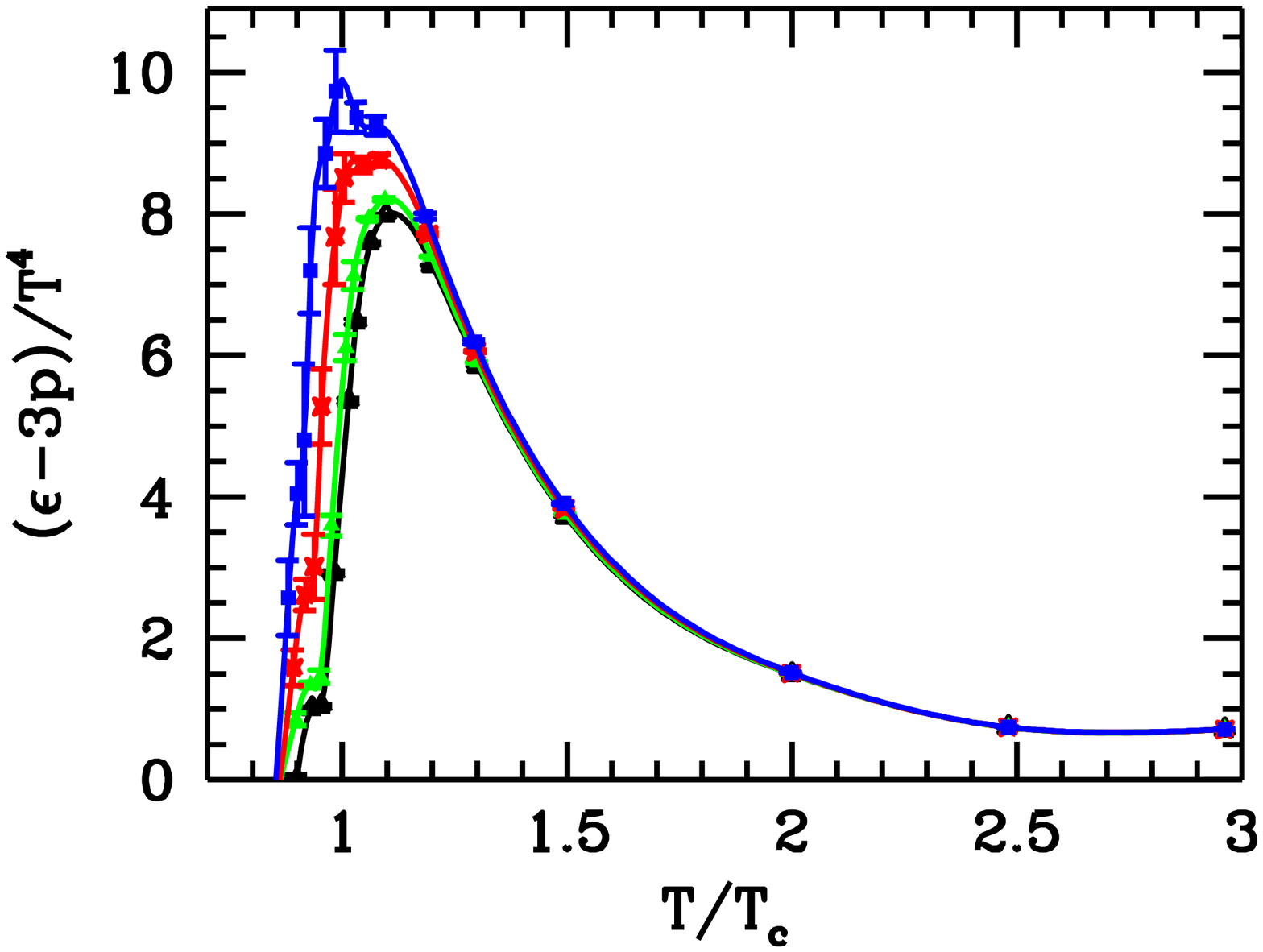,width=82mm}
\hspace*{-0.2cm}\epsfig{file=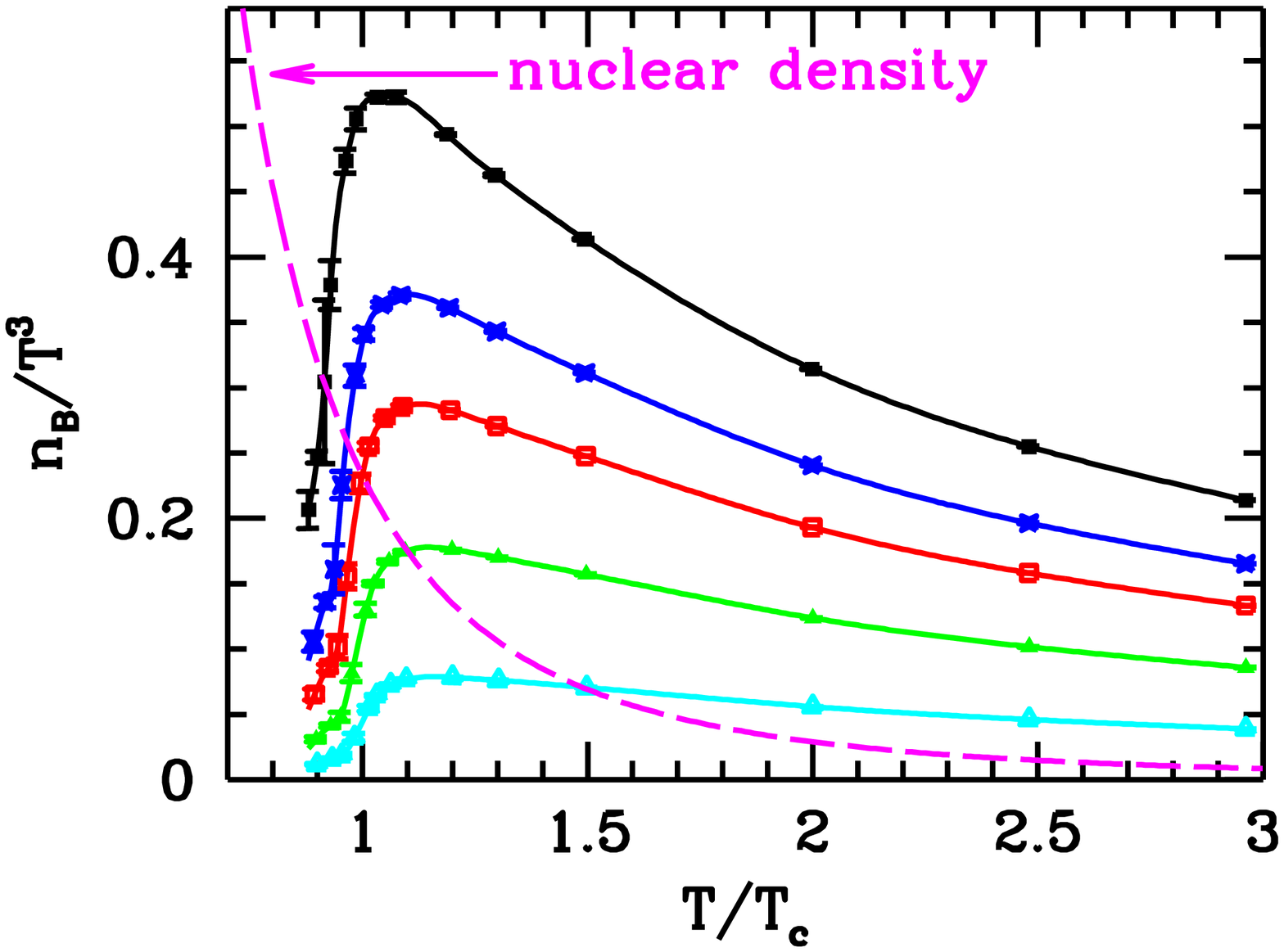,width=82mm}
\vskip -2.9truecm
\caption{
Left panel: 
$(\epsilon-3p)/T^4$ at $\mu_B$=0,210,410~MeV and
$\mu_B$=530~MeV as a function of $T/T_c$
(from bottom to top).
Right panel:
The baryon number density normalized by $T^3$
as a function of $T/T_c$ for $\mu_B=$100,210,330,410~MeV and
$\mu_B$=530~MeV.  As a reference value the line starting 
in the left upper corner
indicates the nuclear density.
}
\vspace{-0.6truecm}
\label{fig:e-3p+n_B}
\end{figure*}

Figure \ref{fig:pressure} shows the pressure ($p$) at $\mu$=0 
and $\Delta p$($T$,$\mu$)=$p$($T$,$\mu$)-$p$($T$,$\mu$=0) normalized by 
$T^4$. Note, that normalizing 
$\Delta p$($T$,$\mu$) by 
$\Delta p$($T\rightarrow\infty$,$\mu$) leads to an almost universal
$\mu$-independent function \cite{Fodor:2002aa}.
The left panel of Figure \ref{fig:e-3p+n_B} 
shows $\epsilon$-3$p$ normalized by
$T^4$, which tends to zero for large $T$ ($\epsilon$ is the energy density). 
The right panel 
gives the baryon number density
as a function of $T/T_c$ for different $\mu$-s. As it can be seen
the densities exceed the nuclear density
by up to an order of magnitude.

\section{Overlap improving multi-parameter reweighting:  
Taylor expansion}

\begin{figure*}[htb]
\vskip -0.3truecm
\vspace{9pt}
\hspace*{0.6cm}\epsfig{file=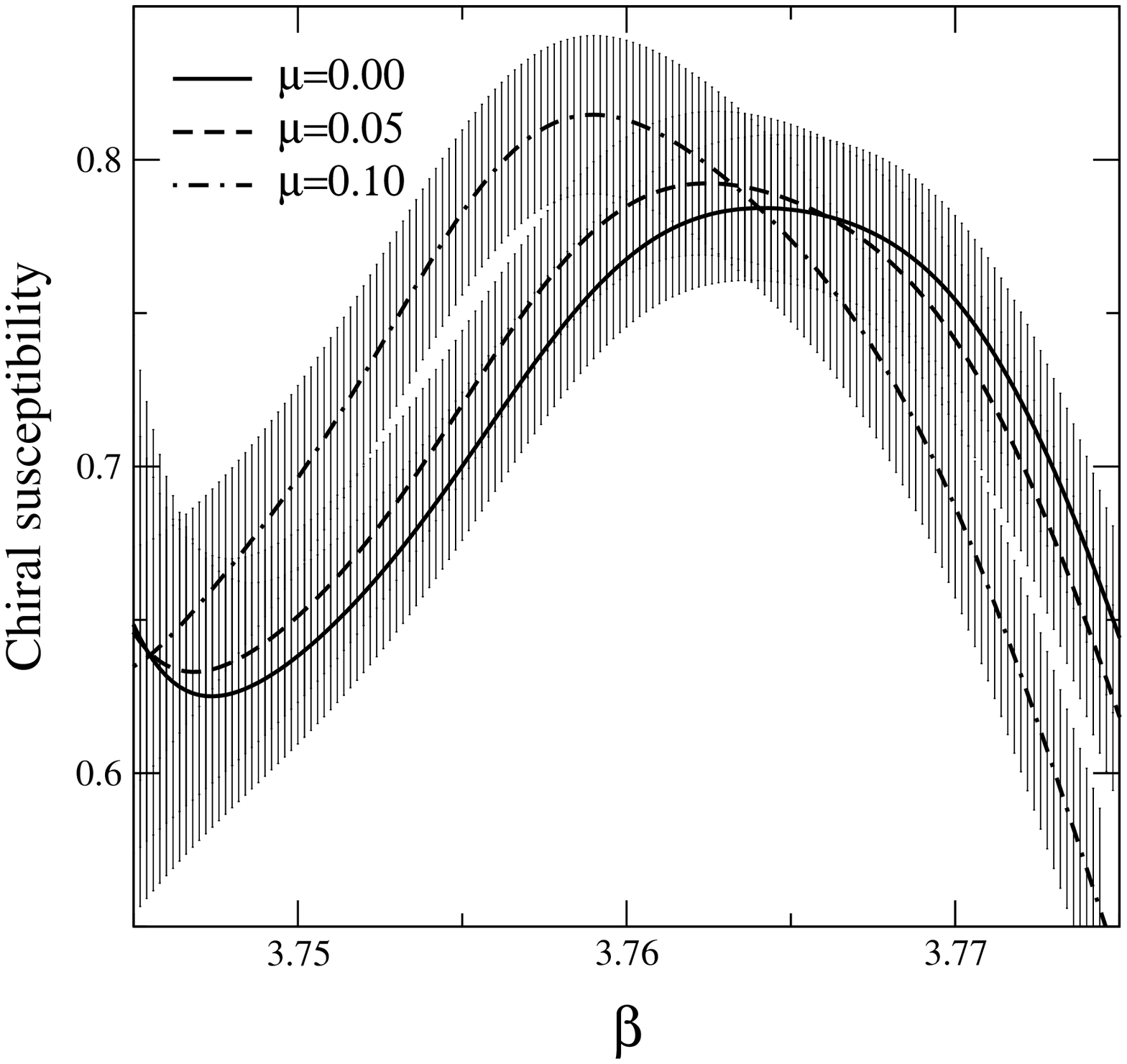,width=64mm}
\hspace*{2.2cm}\epsfig{file=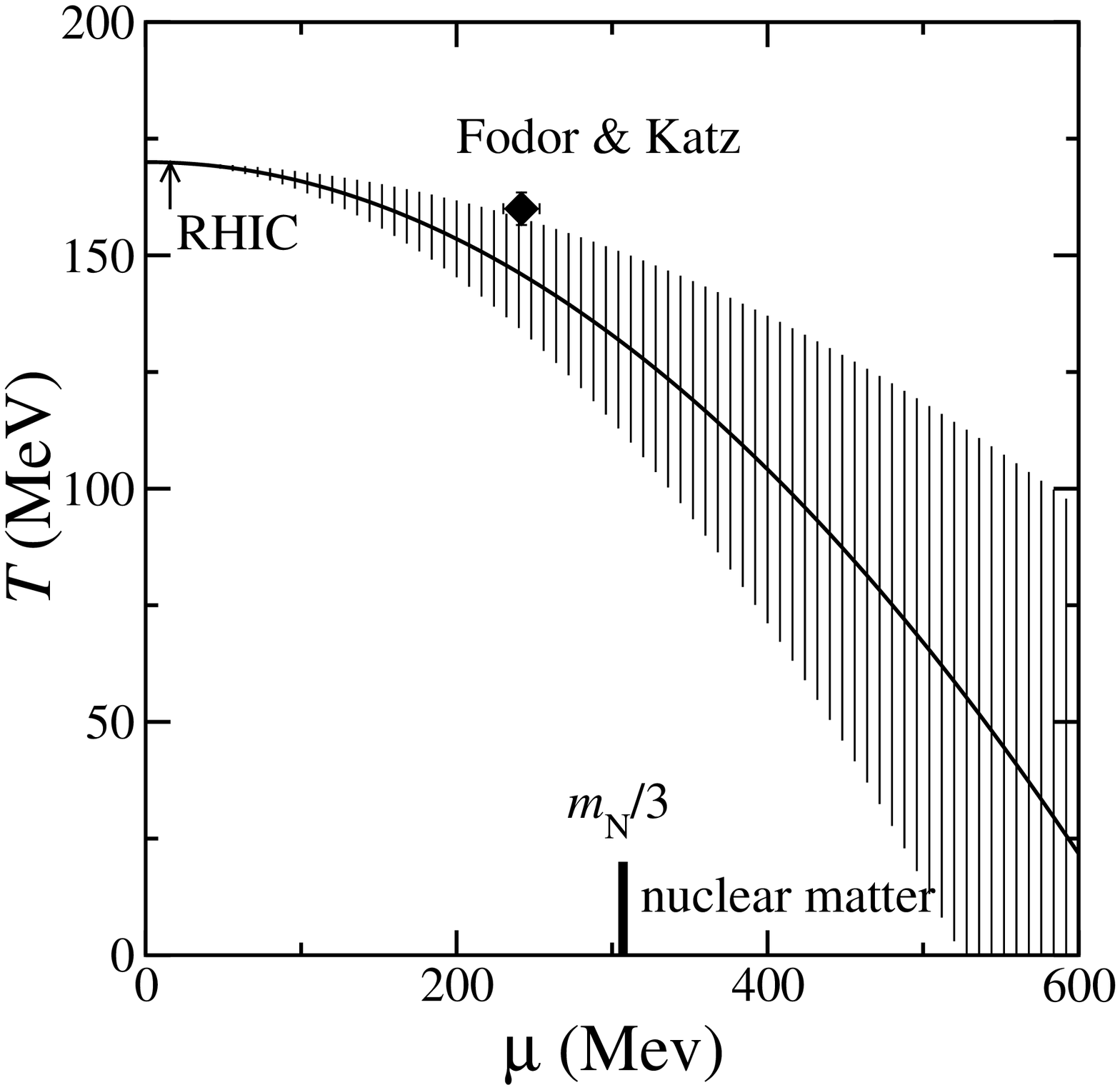,width=62mm}
\vskip -0.9truecm
\caption{
Left panel: The $\mu$ dependence of the chiral susceptibility. 
Right panel:
Transition temperature as a function of $\mu$. The $\mu$ value of RHIC 
is indicated by an arrow.
}
\vspace{-0.6truecm}
\label{fig:taylor}
\end{figure*}

This section summarizes some results of the Bielefeld-Swansea group
\cite{Allton:2002zi,Schmidt:2002aa} obtained 
in dynamical staggered QCD with p4 action on $N_t$=4 lattices. 

Instead of evaluating the determinants in eq.~(\ref{eq:reweight}) 
explicitely one can approximate them by using a Taylor
expansion as given by eq.~(\ref{eq:expand}). One  
calculates observables  
as a function of $\mu$ using eq.~(\ref{eq:opexp}). The left panel of
Figure
\ref{fig:taylor} shows the chiral susceptibility as a function
of the gauge coupling 
for three different $\mu$ values. Note, that the
peak indicates the transition and that its position  
moves as $\mu$ changes. Since the gauge
coupling is directly connected to T one can convert 
the position of the peaks into physical units and obtain the
$T_c$ as a function of $\mu$ 
(right panel of Figure \ref{fig:taylor}). The authors of
Ref.~\cite{Allton:2002zi} concludes that the curvature of the 
phase diagram at $\mu$=0 is in good agreement with that of 
Ref.~\cite{Fodor:2001pe} (they show the endpoint of 
Ref.~\cite{Fodor:2001pe} by 
a diamond signaled as Fodor \& Katz).  

\begin{figure}[htb]
\vspace{-0.9truecm}
\begin{center}
{\includegraphics*[width=16.cm,bb=-0 90 640 320]{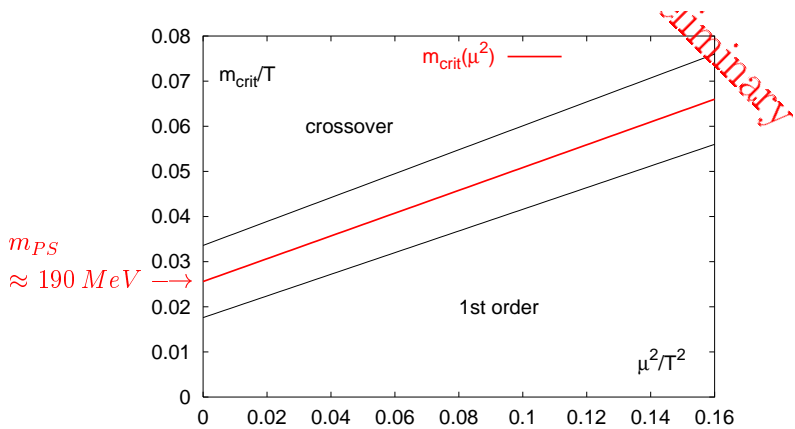}}
\end{center}
\vskip -1.6truecm
\caption{$m_{crit}/T$ as a function of $\mu^2/T^2$. The band indicates the 
uncertainties.
}
\vskip -0.8truecm
\label{fig:endpoint}
\end{figure}

The equation of state was also studied at small $\mu$ values.
At the RHIC point both $\epsilon$ and $p$ increase
by around 1\% from its $\mu$=0 value. Along the transition
line their change is consistent with zero within the current 
precision \cite{Allton:2002zi}.  

In a hypothetical QCD with degenerate quark masses and 
$m_\pi$$\approx$$ m_K$$\approx$190~MeV the critical endpoint is at
$\mu$=0. Note, that $\mu$$\neq$0 values correspond to larger
critical masses ($m_{crit}$). 
A preliminary result is shown in
Figure \ref{fig:endpoint}. Using the $\mu$ dependence
of $m_{crit}$ one could, in principle, estimate 
the location of the endpoint for physical quark masses.

\section{Analytic continuation from imaginary $\mu$}

This section summarizes the 
results obtained by the analytic continuation
technique. $N_t$=4 lattices are used with 2 flavour
\cite{deForcrand:2002ci} and 4 flavour \cite{DElia:2002aa}
dynamical staggered QCD. 

\begin{figure}[htb]
\vskip -0.9cm
\vspace{9pt}
\hspace*{-0.2cm}
{\includegraphics*[width=75mm,bb=10 20 720 540]{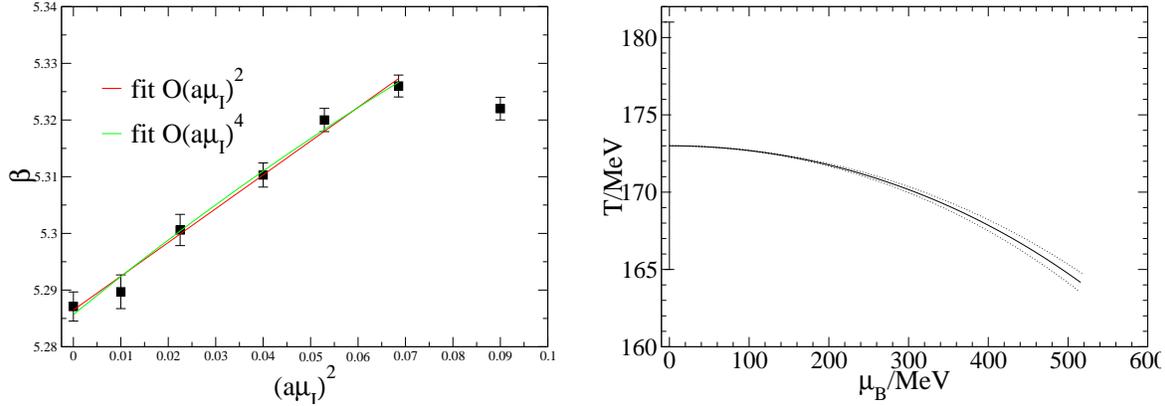}} 
\vspace{-.12cm}
\hspace*{0.2cm}
{\includegraphics*[width=75mm,bb=10 20 720 540]{owe_diag1.eps}} 
\vskip -0.9truecm
\caption{
Left panel: Transition coupling for imaginary $\mu$ with different fits.
Right panel: Transition temperature in physical units for $\mu$$\neq$0.
}
\vspace{-0.8truecm}
\label{fig:immu}
\end{figure}

In Ref. \cite{deForcrand:2002ci} an alternative approach was developed, 
avoiding reweighting in $\mu$ altogether.
This is achieved by simulating with imaginary $\mu$,
where there is no sign problem and hence no need for reweighting.
In this case
one may fit the nonperturbative data
of an observable (even of the critical line)
by truncated Taylor series in $\mu/T$ (see left panel of 
Figure~\ref{fig:immu}).
In the absence of nonanalyticities,
the series may be analytically continued to real values of $\mu$
(right panel of Figure~\ref{fig:immu}).
For the transition temperature \cite{deForcrand:2002ci}
one gets $T_c(\mu)=T_c(0)-0.0056\mu^2/T$ (note, that similar value
was obtained in four flavour QCD by Ref. \cite{DElia:2002aa}). These
results are in complete agreement with those of the previous two 
sections.
 
Performing the analytic continuation for different volumes,
a finite volume scaling analysis should reveal the nature of the
transition.
In particular, it might then be possible to locate
the critical endpoint by this technique, too. 
 
\section{Conclusions}

Due to the notorious sign 
problem this is the first time that 
basic thermodynamic quantities could be obtained on 
the lattice at finite chemical potentials. Completely different
new techniques lead to very similar result, which indicates
that in the near future lattice QCD could be a major
contributor to the field. Clearly, more work has to be done
in order to reach the thermodynamic, chiral and continuum limits.  

\medskip \noindent
{\bf Acknowledgements:} 
The work has been partially supported by Hungarian Scientific
grants, OTKA-T34980/\-T29803/\-M37071/\-OMFB1548/\-OMMU-708. 
Preliminary results from the Bielefeld-Swansea group are also acknowledged
\cite{Schmidt:2002aa}.

\end{document}